\documentstyle[12pt,epsfig]{article}
\newlength{\dinwidth}
\newlength{\dinmargin}
\newcommand{\ba}{\begin{array}}
\newcommand{\ea}{\end{array}}
\newcommand{\be}{\begin{equation}}
\newcommand{\ee}{\end{equation}}
\newcommand{\bea}{\begin{eqnarray}}
\newcommand{\eea}{\end{eqnarray}}

\def\a{\alpha}

\def\d{\delta}

\newcommand{\ep}{\epsilon}

\newcommand{\Om}{\Omega}

\newcommand{\si}{\sigma}

\def\cL{\mbox{${\cal L}$}}

\def\det{{\rm det}}

\def\Vol{{\rm Vol}}

\newcommand{\non}{\nonumber\\}

\newcommand{\p}{\partial}

\def\half{{\mbox{\small  $\frac{1}{2}$}}}

\newcommand{\twomat}[4]{\left( \begin{array}{cc} #1&#2\\ 
#3&#4\end{array}\right)}

\newcommand{\lb}{\left(}
\newcommand{\rb}{\right)}
\renewcommand{\le}{\left\lbrack}
\newcommand{\re}{\right\rbrack}

\newcommand{\beq}{\begin{equation}}
\newcommand{\eeq}{\end{equation}}
\newcommand{\beqa}{\begin{eqnarray}}
\newcommand{\eeqa}{\end{eqnarray}}
\newcommand{\ben}{\begin{enumerate}}
\newcommand{\een}{\end{enumerate}}
\newcommand{\bit}{\begin{itemize}}
\newcommand{\eit}{\end{itemize}}

\newcommand{\refeq}[1]{(\ref{#1})}

\def\adss{$AdS_3\times S^3\;${}}
\def\ads#1{$AdS_#1${}}
\def\S#1{$S^#1${}}
\def\adsss{$AdS_5\times S^5\;${}}

\def\vth{\vartheta}
\def\vthu{\underline \vth}
\def\Fu{\underline F}

\def\cR{{\cal R}}

\begin{document}
\thispagestyle{empty}
\addtocounter{page}{-1}
\begin{flushright}
SNUST-000101\\
{\tt hep-th/0007154}
\end{flushright}
\vspace*{1.1cm}
\centerline{\Large \bf Ramond-Ramond Flux Stabilization of D-Branes~\footnote{
The work of S.-J.R. was 
supported in part by BK-21 Initiative in Physics (SNU 
Project-2), KRF International Collaboration Grant, KOSEF 
Interdisciplinary Research Grant 98-07-02-07-01-5, and KOSEF Leading 
Scientist Program 2000-1-11200-001-1. The work of J.P. was
supported in part by the EC Contract HPRN-CT-2000-00152 and the
Alexander-von-Humboldt Foundation. } }
\vspace*{1cm} 
\centerline{\bf Jacek Pawe{\l}czyk${}^{a,c,d}$ {\rm and} Soo-Jong Rey${}^{b,d}$}
\vspace*{0.6cm}
\centerline{\it Institute of Theoretical Physics, Warsaw University}
\vspace*{0.1cm}
\centerline{\it Hoza 69, PL-00-681, Warsaw, Poland ${}^a$}
\vspace*{0.3cm}
\centerline{\it 
School of Physics \& Center for Theoretical Physics}
\vspace*{0.1cm}
\centerline{\it Seoul National University, Seoul 151-742 Korea ${}^b$}
\vspace*{0.3cm}
\centerline{\it Institut f\"ur Theoretische Physik, Universit\"at M\"unchen}
\vspace*{0.1cm}
\centerline{\it Theresienstr. 37, M\"unchen, Germany ${}^c$}
\vspace*{0.3cm}
\centerline{\it Erwin Schr\"odigger Institute for Mathematical Physics}
\vspace*{0.1cm}
\centerline{\it Boltzmanngasse 9, Wien A-1090 Austria ${}^d$ }
\vspace{0.4cm}
\centerline{\tt jacek.pawelczyk@physik.uni-muenchen.de $\hskip1cm$
sjrey@gravity.snu.ac.kr}
\vspace*{1cm}
\centerline{\bf abstract}
\vspace*{0.4cm}
In $AdS_n \times S^m$ threaded by $N$-units of Ramond-Ramond flux, the 
dynamics of a D$m$-brane wrapped partially on $S_{m-1} \subset S_m$ is 
investigated. Under the condition that flux of the {\sl dual} gauge field 
on the D-brane worldvolume is quantized integrally, it is found that the 
wrapped D-brane is stable both locally and globally and, on $AdS_n$, 
behaves effectively as a fundamental string. It is also claimed that the 
semi-infinite, partially wrapped D$m$-brane can `lasso' around the $S^m$.
\vspace*{1cm}

\baselineskip=18pt
\newpage

\section{Introduction}
Recently, it has been shown that a D2-brane wrapped on $S_2 \subset S_3$
is a stable configuration \cite{jbds}, despite that $H_2 (S_3) = 0$. 
The stability is provided by a nontrivial NS-NS background potential, threaded 
through $S_2$. The argument of \cite{jbds} relied heavily on the CFT 
description of open strings in NS-NS background in terms of SU(2) WZW model
\cite{as}.

In this paper, we would like to explore further D-brane dynamics in a
nontrivial background, in particular, in the R-R background. While some
of the examples are related to the NS-NS background by S-duality, we will 
also uncover several new features as well as certain puzzling aspects, 
e.g. fate of the charges of the effective branes extending in 
$AdS$ part of the background. We shall analyze the dynamics of branes in
terms of the DBI action. Thus, most of the calculations would be similar
to those of \cite{jbds}. However, there is one important difference --- 
for instance, in $AdS_3 \times S^3$ space, one could have disregarded the 
$AdS$ part of the dynamics for NS-NS background, but not for R-R background.

The set-up of D-brane configurations considered in this paper will be 
as follows. We shall be looking for a classical configuration of D$m$-brane
embedded in $AdS_n \times S^m$, in which flux of R-R tensor field is 
threaded through $S^m$. In particular, we shall consider D$m$-branes 
wrapped on $S^{m - 1} \subset S^m$ ($m$ = 3, 5) and show that they transmute
into effective F-strings extended into the $AdS$ space. In section 2 and 3, 
we will study D3-brane in $AdS_3 \times S^3$ space and D5-brane in $AdS_5 
\times S^5$ space, respectively. In section 4, we will discuss charge
nonconservation and relation to baryon vertex. We conclude, in section 5,
with posing several unresolved, open problems.  
    
\section{D3-brane in \adss(R-R)}
\subsection{Brane Configuration}
We begin with classical configuration of a D3-brane in \adss. String theory 
in this background has been studied extensively in numerous recent works 
\cite{mald-strom,d3rr}. The relevant supergravity background in the near 
horizon limit is 
\beqa
ds^2 &=& \ell^2 \a' \le u^2(- dt^2 +(dx)^2) +   
\frac{du^2}{u^2} + d\Omega_3^2\re\non
H^{\rm RR}&=&2 Q_5 \a' (\ep_3+*_6\ep_3)\non 
e^{-2\phi }& =& {1 \over g_6^2} {Q_5 \over Q_1},\quad
\nonumber
\eeqa
where $*_6$ denotes Hodge duality in $AdS_3\times S^3$ space, $\ep_3$ is the
volume element of the unit 3-sphere, and $\ell^2 = g_6 \sqrt{Q_1 Q_5}$. 
The background is characterized by 
the constant dilaton and the non-trivial R-R 3-form fields. We will find it
convenient to represent the metric on $S^3$ as  
$d\Om_3^2=d\vth^2+\sin^2\vth d\Om_2^2$: a 2-sphere $S^2$ of radius 
$\sin\vth$ is located at latitude angle $\vth$.

At leading order in $\a'$, D3-brane world-volume effective action in the above 
background is given by
\beq
S_{\rm DBI}=T_3 \lb -\int_{\Vol}d^4\si
  e^{-\phi}\sqrt{-\det[(X^*G+2\pi\a' F)_{ab}]}
\pm 2\pi\a'F\wedge X^*C_2^{\rm RR} \rb .
\nonumber
\eeq
If we rescale as $F \rightarrow \ell^2 F$ and factor out all powers of 
$\a', \ell^2, Q_5$ from the metric and $C_2^{\rm RR}$, then the above action
becomes
\beq\label{d3action}
S_{\rm DBI} = T_3 Q_5 \ell^2 \a'^2 \lb
-\int_{\Vol} d^4 \si \sqrt{ -\det[(X^*G + 2 \pi \a' F)_{ab} ]}
\pm 2 \pi F \wedge X^* C_2^{\rm RR} \rb.
\eeq
Locally, we may integrate $H^{\rm RR}=dC_2^{\rm RR} $ and obtain the 2-form 
potential $C_2^{\rm RR}$ on $S^3$: $C_2^{\rm RR} = 
(\vth-\nu-\half\sin(2\vth))\,\ep_2$, where
$\nu$ is an integration constant and $\ep_2$ denotes the volume-form of the 
unit $S^2$.  It is not possible to define $C_2^{\rm RR}$ globally over the
whole \S3, as $H^{\rm RR}$ is a non-trivial element of $H^3(S^3,R) = {\bf Z}$.

Consider now a D3-brane embedded in \adss, whose world-volume is extended 
along 
$\si^\mu=(t,u,S^2)$ and is located at fixed $\vth=\vthu$ and $x$. 
The embedding implies that the pull-back of the two-form R-R potential
$X^*C_2^{\rm RR}$ has components only along $S^2$ in \S3. Hence, among the
world-volume gauge field strength, only the $F_{0u}$ component couples linearly
to the background R-R potential $C_2^{\rm RR}$ through the Chern-Simons term.
The coupling in turn leads to $F_{0u}\neq 0$, which is needed  for a 
non-trivial extremum to exist. As the partially wrapped D3-brane is extended
along $u$-direction, the coupling also implies that the D3-brane behaves 
effectively as a fundamental string(F-string). 
We have found two classes of the extremum of \refeq{d3action}:
\beqa\label{null}
2\pi\Fu &=&  {\rm constant}, \quad \qquad \vthu = 0,\pi\\
\label{sph}
2 \pi \Fu&=&\mp \cos\vthu, \quad \qquad \vthu= {\rm constant}.
\eeqa
The first one corresponds to the collapsed cycle and will not be discussed
here. The second family of solutions depends on $\vthu$ as a moduli space
parameter.
In the S-dual situation, where nontrivial background involves NS-NS tensor
field \cite{jbds}, the $\vthu$ moduli space has been discrete as a result
of the requirement that 
the effective D-brane carries an appropriate R-R charge. In the present case,
the effective fundamental string in \ads3 carries NS-NS charge, which can be
derived from terms in \refeq{d3action} linear in the NS-NS 2-form $B_2$. 
Taking into account of the non-trivial background of $\Fu$, we find:
\beq\label{3charge}
T_3\int_{S^2} \frac{\d S_{\rm DBI}}{\d (2\pi\a' F_{ab})}\;X^*B_{ab} = 
\mp T_3 Q_5 \a' \Vol(S^2)\,(\vthu-\nu)\;X^*B_{0u} ,
\eeq
where $T_{Dp}=1/((2\pi)^p\a'^{(p+1)/2}g_{st})$ and the last expression has 
been derived at the classical extremum. Recall that 
$\d S_{\rm DBI}/{\d (2\pi F_{ab})}$ is proportional to the {\sl dual} 
gauge field strength ${\ep_{ab}}^{cd} {\tilde F}_{cd}$ 
\cite{dualDBI}. Thus, in a simplified 
notation, the NS-NS electric charge density in \refeq{3charge} has the form 
$T_3 [ X^*B\wedge(2\pi\a' {\tilde F}) ]$, viz. the string charge density 
agrees precisely with an expression which is electric-magnetic {\sl dual} to 
the one obtained for the case \adss with NS-NS 3-form flux \cite{jbds}. 
Indeed, this is what one would have expected from the S-duality transformation
of the Type IIB string background and the self-duality of the D3-brane.
Thus, at strong coupling, \refeq{3charge} should be the correct expression,
as it is precisely S-dual to the known situation with NS-NS 
background \cite{jbds}. In particular, precisely as in \cite{jbds}, the 
integral flux quantization in \refeq{3charge} ought to refer to the 
{\sl dual} gauge field $ (2 \pi \alpha') \widetilde{F}$ itself, not to the 
one involving the R-R two-form: $(2 \pi \alpha') \widetilde{F} - X^* 
C_2^{\rm RR}$. At weak coupling regime, we do not expect corrections to the 
definition of the charge \refeq{3charge}, as the wrapped D-brane is a BPS 
state. 

We can also fix the integration constant $\nu$ by demanding that the NS-NS 
charge  should be zero for $\vthu=0$. This sets $\nu=0$.
The flux quantization condition implies that the last expression of 
\refeq{3charge} equals the {\sl integer} multiple of the fundamental string 
charge. It yields 
\beq
\vthu\,{}_n=\frac{n}{Q_5}\pi ,\quad \qquad 0 \leq n \leq Q_5 .
\eeq 
This is precisely the same result as for the \adss with nontrivial 
NS-NS background. However, in contrast to the NS-NS background case, we do not 
have yet any direct argument for the above quantisation from boundary conformal
field theory or other string theoretic framework.
One can also derive the tension of the effective string as:
\beq\label{3mass}
{\cal E} \, = \, \left( F_{0u}\frac{\d\cL}{\d F_{0u}}-\cL \right) 
= T_3 Q_5 l^2 \a'^2 \Vol(S^2)\,\sin\vthu.
\eeq

The results \refeq{3charge} and \refeq{3mass}, which are main results of this
section, may be summarized by posing the following puzzle. The NS-NS charge 
density \refeq{3charge} does not exhibit $\vthu\to \pi-\vthu$ symmetry, whose 
origin can be traced back to the fact that the R-R 2-form potential 
$C_2^{\rm RR} $ cannot be defined globally, as $H_3(S^3) = {\bf Z}$. 
On the other hand, the tension is a periodic function of $\vthu$, being
proportional to $\sin\vthu$. Thus, it appears that the charge density and 
the tension are not proportional to each other, except near the north pole,
$\vthu \sim 0$. Moreover, comparison of \refeq{3charge} and \refeq{3mass} 
suggests that the charge of the effective string is defined only modulo
a multiple of $Q_5$ (at large $Q_5$ approximation). These are apparently 
the same phenomena as for the NS-NS background and calls for a deeper 
understanding.  
\subsection{Stability}
The D3-brane is wrapped partially on $S^2 \subset S^3$, hence, is apt to
shrink down to a singular configuration. Let us begin with local analysis of the brane system.
Denote the spacetime coordinates as $(012)(345)(6789)$, where
$(012)$ is along the $AdS_3$ directions, $(345)$ is along the $S^3$
directions, and $(6789)$ are the four-dimensional spectator directions.
The \adss is provided by stack of $D1$ and $D5$ branes whose world-volumes
are oriented along (01) and (016789) directions, respectively.
The partially wrapped D3-brane world-volume is oriented along (0234) directions.
As such, the D3-brane is a supersymmetric configuration  
with respect to both the D1-branes with relative co-dimension 4 and
the D5-branes, with relative co-dimension 8. 
Hence, the partially wrapped D3-brane ought to be a stable configuration,
at least locally for each point on $S^2 \subset S^3$. 

We next examine global stability of the classical solution explicitly by 
expanding D3-brane world-volume fields in harmonic fluctuations around the 
classical configuration.
Denote small fluctuations of the $\vth$-field by $\xi$ and those of 
$F$-field by $f$. 
From \refeq{d3action}, we get the following effective string Lagrangian density
for $\xi,\,A_u$ up to quadratic terms 
(after rescaling $A_u\to\sin\vthu\, A_u$)
:
\beqa
\cL={T_3 \over 2} 
\sin\vthu \int_{S^2} \left[
{1 \over u^2} (\p_0 \xi)^2 - u^2 (\p_u \xi)^2-(\p_i \xi)^2
+2\xi^2  + f^2\mp 4 f \xi -u^2\,F_{iu}^2 \right] .
\eeqa
Here, the indices $i$ refers to the coordinates on $S^2$.
Choosing the gauge $A_0=0$ so that $f=\p_0A_u$, the Gauss' law constraint reads
\beq\label{constr}
0=\frac{\d \cL}{\d A_0}=\p_u (\p_0A_u\mp 2 \xi) .
\eeq
Let us expand the fluctuating fields into spherical harmonics on $S^2$:
\beqa
\xi(t, u, {\bf y}) =\sum_{l \ge 0}  Y_{(l,m)} ({\bf y}) \xi_{(l,m)} (t,u),
\quad 
A_u(t, u, {\bf y}) =\sum_{l>0} Y_{(l,m)} ({\bf y}) \a_{(l,m)} (t, u).
\eeqa
Then, the equations of motion for $\a,\,\xi$ read
\beqa\label{eqm-eff}
&&u^{-2} \p_0^2\xi-\p_u \left(u^2 \p_u \xi \right)+
l(l+1)\xi-2\xi\pm 2\p_0 \a=0 \nonumber \\
&&\p_0^2\a\mp 2\p_0\xi+l(l+1) u^2 \a=0 ,
\eeqa
where we have suppressed the $(l,m)$ indices, as they are common to all the 
fields at the linearized level. We change the field variables by introducing 
a new field $\eta=\p_0\a\mp2\xi$. 
For $l=0$ mode, the second equation in \refeq{eqm-eff} gives 
$\p_0\eta=0$ i.e. $\eta$ is not a
dynamical field. The first equation in \refeq{eqm-eff} 
can be solved substituting $\eta=0$.  It then leads to 
$$
\frac1{u^2}\p_0^2\xi-\p_u(u^2 (\p_u \xi))+2\xi=0
$$
viz. the $l=0$ mode is massive.
Next, let $l>0$, then differentiating the second equation with respect to time 
variable we get 
\beqa
&&u^{-2} \p_0^2\xi-\p_u \left(u^2 \p_u \xi \right)+
[l(l+1)+2]\xi\pm 2\eta=0 \nonumber \\
&&\p_0^2\eta+l(l+1) u^2(\eta\pm2\xi) =0 .
\nonumber
\eeqa
The resulting mass matrix takes the form: 
\beq
\twomat{l(l+1)+2}{\pm2}{\pm2l(l+1)}{l(l+1)}
.
\nonumber
\eeq
It is exactly the same mass matrix as obtained in \cite{jbds}.
Thus, the fluctuation spectrum is
\beqa
m^2_l=\left\{ 
\begin{array}{ll} (l+1)(l+2)  &\qquad \mbox{for } l=0,1,...\\
(l-1)l&\qquad \mbox{for } l=1,2,...  \end{array}\right.
\eeqa
It is clear that there is a triplet of massless modes for $l=1$, corresponding
to the Goldstone modes of the broken rotational 
symmetry $SO(4)\to SO(3)$. 
All higher modes are massive and hence proving the harmonic stability of the 
partially wrapped D3-brane.

This stability analysis shows that the partially wrapped D3-brane has the same 
properties as S-dual counterpart  viz. the case with NS-NS 
background \cite{jbds}. The agreement is quite suggestive because even the 
massive spectra of quadratic fluctuations are the same. 
This might indicate that even the D3-brane wrapped on $S^2$-cycle in R-R 
background can be described in terms of fuzzy geometry \cite{myers}. 


\section{D5-brane in \adsss (R-R)}
\label{d5}
\subsection{Brane Configuration}
We now study dynamics of partially-wrapped D5-brane in \adsss threaded by
R-R flux. The situation is parallel to that of the D3-branes in 
\adss, so we will be brief and limit foregoing discussion only to essential 
steps. The ten-dimensional Type IIB supergravity background is given 
by \cite{hs-duff}
\beqa
ds^2&=& \cR^2 \a'  \left[
u^2 ( - dt^2 + d{\bf x}_\parallel^2 ) + {du^2 \over u^2} 
+d\Omega_5^2\right]\non
G_5&=& 4 \cR^4 \a'^2 (\ep_5+*\ep_5)\\
e^{-\phi}&=&1,\qquad \quad
{\cal R}^2=\sqrt{4\pi g_{\rm st} N} .
\eeqa
Here, $\ep_5$ denotes the volume element of the $S^5$.
Again, consider a D5-brane, wrapped partially on $S^4 \subset S^5$.
The world-volume effective action of the D5-brane is given by
\beq\label{d5action}
S_{DBI}=T_5 \lb -\int_{\Vol}d^4\si
  e^{-\phi}\sqrt{-\det[(X^*G+2\pi\a' F)_{ab}]}\pm 2\pi\a'F\wedge X^*C_4\rb
\eeq
As before, decompose the metric on $S^5$ in polar coordinates so that 
$d\Omega_5^2=d\vth^2+\sin^2\vth\, d\Omega_4^2$, where $\Omega_4$ denotes
the metric on $S^4$, and consider the D5-brane whose world-volume wraps 
partially on $S^4$ at a fixed $\vth$ and also extends along $u$-direction in
\ads5. Due again to the Chern-Simons coupling, the $F_{0u}$ component of the
world-volume gauge field acquires a non-vanishing expectation value,
$F_{0u}=\mp \cos\vthu$.

In $AdS_5$, extended along $u$-direction, the partially wrapped D5-brane
behaves as an effective string.
Again, the effective string is nothing but a multiple of fundamental string.
To see this, we calculate the  string charge density from 
the minimal coupling of the D5-brane to the NS-NS 2-form $B_2$. Taking into
consideration of the non-trivial $F$-field background, from \refeq{d3action},
we find the following effective coupling with $B_2$:
\beqa\label{5charge}
T_5 \int_{S^4}  \frac{\d S_{DBI}}{\d (2\pi\a' F_{ab})}\;X^*B_{ab}
&=& \mp T_5 \cR^4 \Vol(S^4)\,\frac32(\vthu-\half\sin2\vthu) \;X^*B_{0u}
\\
&=& \mp\frac{N}{2\pi^2\a'}(\vthu-\half\sin2\vthu) \;X^*B_{0u} .
\nonumber
\eeqa
Here, the last equality has been calculated at the classical extremum.
As in the case of D3-brane in \adss (R-R) discussed in section 2,
 we did not include in \refeq{5charge} the field corrections proposed 
in \cite{wati}. The charge formula \refeq{5charge} shows that, 
for a given D5-brane, the effective string charge can not exceed $N$. 
  
On the other hand, as in \adss (R-R), the tension of the effective string can 
be estimated from the mass density and turns out to be proportional to  
$\sin^3\vthu$. 
Once again, the charge density and the tension of the effective
string are not proportional to each other. Only around $\vthu=0$, they
both grows as $\vthu^3$ and hence become proportional. 
Thus, we conclude that the properties of the effective 
strings in this case are analogous to the \adss case. 

\subsection{Stability}
Stability analysis of the partially wrapped D5-brane n \adsss proceeds 
essentially as in the \adss case. Local analysis shows 
 that the partially wrapped brane should be a BPS state.
Denoting by (0123) the directions along D3-brane (whose near horizon 
geometry is \adsss), we embed the D5-brane in (045678). As the relative 
co-dimention is 8, the brane system preserve half of the supersymmetries 
of the background. 

We next proceed with the global analysis of harmonic fluctuations.
Expanding the fluctuations of the $\vth$ and $A_u$ fields in spherical
harmonics on $S^4$, we find: 
\beqa
&&u^{-2} \p_0^2\xi-\p_u \left(u^2 \p_u \xi \right)+
l(l+3)\xi-4\xi\pm 4\p_0 \a=0 \nonumber \\
&&\p_0^2\a\mp 4\p_0\xi+l(l+3) u^2 \a=0 .
\nonumber
\eeqa
As in the previous section, we introduce a new field $\eta=\p_0\a\mp4\xi$. 
Then, the mass matrix for $\xi,\,\eta$ ($l>0$) is given by
\beq
\twomat{l(l+3)+4}{\pm4}{\pm4l(l+3)}{l(l+3)}
,
\nonumber
\eeq
yielding the fluctuation spectrum as 
\beqa
m^2_l=\left\{ 
\begin{array}{ll} (l+3)(l+4) &\quad \mbox{for } l=0,1,...\\
(l-1)l &\quad \mbox{for } l=1,2,...  \end{array}\right.
\nonumber
\eeqa
For $l=1$, we get 5 Goldstone modes in fundamental representation of $SO(5)$ 
arising from spontaneously broken rotational symmetry $SO(6)\to SO(5)$. 
As anticipated from vector spherical harmonics, the above spectrum bears 
a similar structure as that in the \adss case.

\section{Charge Nonconservation and Baryon Vertex}
In the previous sections, we have studied partially wrapped D3- or
D5-brane on $S^2 \subset S^3$, respectively, $S^4 \subset S^5$.
On $AdS_3$ or $AdS_5$ space, the brane was assumed to extend along
the $u$-direction infinitely $u = [0, \infty]$, 
hence, connecting two boundary points  of the anti-de Sitter space. 

Now, let us consider a variant situation: 
the partially wrapped D-brane
forms a {\sl semi-infinite} string running between $u = [ u_0, \infty]$
with $N$ units of NS-NS charge\footnote{We set $N=Q_5$ in this section.}.
The configuration is clearly unstable. For one thing, due to nonzero
tension, the effective string will shrink to a zero size. In this case,
we will need to understand one more piece of physics: fate of the string 
charge. We now would like to argue that, instead of shrinking to the
boundary of anti-de Sitter space, the semi-infinite effective string, 
whose configuration on $S^3$ or $S^5$ is a co-dimension one hypersphere, 
will `lasso' around $S^3$ or $S^5$ completely, say, from the North- to 
the South-pole. If one assumes the mod-$N$ charge conservation rule, 
then we find that the same mechanism will stabilize a single string.
>From the charge density formula, we note that the 
charge deficiency when `lassoing' around the sphere is precisely
$N$ units, which is the same amount as the induced effective string charge. 

The final configuration would then be ( for \adss) 
the superposition of the D7-branes:
a D7-brane wrapped partially on $S^2 \subset S^3 $ and its `lasso' image 
wrapped entirely on $S^3$.
The superposed D7-brane composite is a configuration of relative co-dimension
two, hence, corresponds to a non-threshold bound-state. Indeed, the bound-state
is formed precisely by the world-volume gauge field, which is directly coupled
to the NS-NS two form.    
Thus, the above argument provides a complementary picture of the baryon vertex
from the string creation phenomenon. Namely, the partially wrapped D-brane 
will try to `lasso' around the entire $S^m$ and absorbs the induced NS-NS
fluxes on $S^m$. 

In order to see whether this is the case, let us reconsider the baryon 
vertex in \adsss. 
>From the ${\cal N}=4$ super Yang-Mills theory point of view, the baryon
is constructed by wrapping a D5-brane on $S^5$ at $u = u_0$ and  connecting 
$N$ fundamental string between the wrapped D5-brane and $N$ D3-branes at 
the boundary of $AdS_5$. 
World-volume dynamics of the D5-brane wrapped on $S^5$ exhibits two novel 
features. First, the D5-brane carries $N$ units of fundamental string charge. 
This is because world-volume action of the D5-brane wrapped on $S^5$ 
includes the following Wess-Zumino term \cite{w-bar}:
\bea
S_{\rm WZ} &=& \int_{S^5 \times T} (B + 2 \pi \alpha' F) \wedge C_4
\nonumber \\
&=& - \int_T (\Lambda_1 + 2 \pi \alpha' A) \wedge \int_{S^5} G_5.
\eea 
Utilizing that $\oint_{S^5} G_5/ 2\pi = N$,    
we conclude that $N$ units of electric flux or, gauge equivalently, 
$N$ units of fundamental string charge is induced on the D5-brane. 
As the world-volume of the D5-brane wrapped on $S^5$ is compact, 
Gauss' law dictates that the flux has to leak out to `infinity'. 
This is achieved by attaching the $N$ fundamental strings connecting
the D5-brane wrapped on $S^5$ and the D3-branes -- the flux on D5-brane
then leaks out to the D3-branes (whose world-volume is noncompact) via
the $N$ fundamental strings. Now, the requisite $N$ fundamental strings are 
provided precisely by the D5-brane wrapped on $S^4 \subset S^5$. In 
other words, the entire configuration of the baryon vertex and the
fundamental string is made solely out of two intersecting D5-branes -- one
wrapped on $S^5$ and another wrapped on $S^4 \subset S^5$!  
Second, the D5- and D3-branes have relative co-dimension $8$, so form a
supersymmetric configuration. We have seen that the D5-brane wrapped partially
on $S^4 \subset S^5$, which behaves effectively as $N$ fundamental strings
stretched along $u$-direction, also have relative co-dimension $8$, so is a 
supersymmetric configuration. In doing so, the aforementioned two D5-branes 
wrapped on $S^4$ and $S^5$, respectively, form a non-threshold bound-state.  
This is exactly the same situation as the system of D7- and D3-branes exhibits
in \adss, as already mentioned.

\section{Open problems}
Our discussion is far from conclusive and calls for deeper understanding of 
the physics of branes on $AdS_m \times S^n$ spaces. Let us list here several 
open problems. It is clear that our results are subject to
curvature corrections \cite{bbg}, as the D-branes are wrapped on $S^{n-1}
\subset S^n$. What is more important, we lack a proper definition of the
D-brane charges when various background fields are present. Our calculations
suggest that the charge might be conserved modulo $N$ ($Q_5$ for \adss) (see
also \cite{as-charge}), but we do not know any possible mechanism underlying 
such a behaviour directly from the D-brane physics.  We also lack a proper
understanding of corrections to effective string charges of the type 
described in \cite{wati} in the R-R field background. The S-duality enforces 
lack of such contributions for the case of the  R-R background, while they 
were necessary for the NS-NS background. Lastly, the partially wrapped 
D-branes have unusual charge-tension relations, which certainly calls 
for further studies.

\vskip1cm 
\section*{Acknowledgement}
We thank A. Recknagel, S.Ramgoolam and S. Theisen for interesting conversations.
We are grateful to organizers, H. Grosse, M. Kreuzer
and S. Theisen, for 
inviting us to "Duality, String Theory and M-Theory" workshop at Erwin
Schr\"odinger Institute, where part of this work has been accomplished.

\end{document}